\documentstyle[aps,epsf,prl,preprint]{revtex}
\begin{document}

\title{The Spatio-Temporal Structure of Spiral-Defect Chaos}

\author{Stephen W. Morris}
\address{Department of Physics and Erindale College, University of Toronto}
\address{60 St. George St., Toronto, Ontario, Canada M5S 1A7}

\author{Eberhard Bodenschatz}
\address{Laboratory of Atomic and Solid State Physics, Cornell University}
\address{Ithaca, New York, 14853.}

\author{David S. Cannell and Guenter Ahlers}
\address{Department of Physics and Center for Nonlinear Science}
\address{University of California, Santa Barbara, CA, 93106-9530}

\date{\today}
\maketitle
\begin{abstract}
We present a study of the recently discovered spatially-extended chaotic state known as {\it spiral-defect chaos}, which occurs in low-Prandtl-number, large-aspect-ratio Rayleigh-B{\'e}nard convection.  We employ the modulus squared of the space-time Fourier transform of time series of two-dimensional shadowgraph images to construct the structure factor ${S}({\vec k},\omega )$.   This analysis is used to characterize the average spatial and temporal scales of the chaotic state.  We find that the correlation length and time can be described by power-law dependences on the reduced Rayleigh number ${\epsilon}$. These power laws have as yet no theoretical explanation.

\end{abstract}


\section{Introduction}
\label{intro}

Attempts to understand the complex deterministic motion of spatially extended nonlinear dissipative systems, so-called {\it spatio-temporal chaos} (STC), are at the forefront of research in nonlinear dynamics\cite{CH92,CH94}.  In contrast to simple chaotic systems in the time domain, in which a few important degrees of freedom are nonlinearly coupled, spatially-extended systems seem to require an infinite number of modes for their description.  Simple chaotic systems are typically described by a few amplitudes which are determined by coupled nonlinear ordinary differential equations. Extended systems, on the other hand, must be described by nonlinear {\it partial} differential equations for which, in general, no natural reduction to a few modes is known. This extended form of chaos has been characterized to some extent for a variety of experiments and simulations, but continues to resist any understanding based on generalizations of the few-modes case.  In this paper, we describe experiments on the spiral-defect-chaos (SDC) state\cite{MBCA93,HEA93,AS94,HEA95a,HEA95b,EHMA95} in Rayleigh-B{\'e}nard convection\cite{CH92,RBC,Ma90} (RBC). RBC occurs beyond the first hydrodynamic instability of a thin horizontal fluid layer heated from below, and is particularly suitable for the study of complex pattern-formation phenomena because so much of the theoretical foundation has been established by the extensive stability analyses of Busse and coworkers.\cite{SLB65,Bu67,CB74,BC79}

The SDC state consists of a disordered pattern of convection rolls, which prominently features the persistent spontaneous appearance and disappearance of rotating spiral defects.  It offers an unusually controllable experimental system for the study of STC.  Other examples of STC which have been studied experimentally include for instance ``dispersive chaos" in binary fluid convection\cite{KGW90}, chaotic regimes in electro-convection in nematic liquid crystals\cite{NK91,BRS91,DAC95},  chemical-reaction patterns\cite{VOS92}, parametrically-excited surface waves\cite{TRG89}, and RBC under rotation in the K{\"{u}}ppers-Lortz unstable regime\cite{HEA95c}.  Most of these cases have the common feature that STC is encountered following the breakdown of an ordered pattern as a control parameter is increased.  The onset of STC in such systems is typically linked to the appearance and proliferation of defects in an underlying pattern, leading to complex, persistent pattern dynamics.  These patterns, although disordered, still have a characteristic wavenumber and are far simpler than fully developed turbulence in which a broad spectrum of length scales is involved.  Usually, a crucial role is played by the size of the system, which must be large when measured in units of the characteristic size of the pattern elements.  Only in the limit of large system size, may one hope to approach the limit of a statistically homogeneous chaotic state that is free of boundary effects.

RBC may be parameterized by four dimensionless numbers.  The Rayleigh number ${R}$ is a dimensionless temperature difference across the layer and is given by 
\begin{equation}
{\it R} = {{g\alpha d^3\Delta T }\over{\kappa\nu}}\ .
\end{equation}
It has a critical value of ${R_c~=~1708}$ for a laterally infinite layer at the onset of convection. Here ${\Delta T}$ is the temperature difference, $d$ the layer thickness,   $g$ the acceleration due to gravity, $\alpha$ the thermal expansion coefficient, $\kappa$ the thermal diffusivity and $\nu$ the kinematic viscosity.   It is convenient to define the reduced Rayleigh number ${\epsilon~\equiv~(R/R_c)-1}$, so that ${\epsilon~=~0}$ corresponds to the onset of convection.  This number may be regarded as the experimental control parameter.  The Prandtl number, given by ${{\sigma} = {\nu}/{\kappa}}$, is a second parameter that describes the fluid properties.  It has important effects on what secondary instabilities are expected in the pattern that is found above onset\cite{CB74,BC79}.  In our experiments, the working fluid was pressurized ${CO_2}$ gas, for which ${{\sigma} \approx 1}$. This value of $\sigma$ 
is characteristic of gases.  A Prandtl number close to or less than one seems crucial for the appearance of SDC. The lateral extent of the convecting layer is described by the aspect ratio ${{\Gamma} = r/d}$, where $r$ is taken to be the radius for a circular geometry.  The SDC state is found only for ${\Gamma \gg 1}$.  The lateral boundaries may also influence the pattern {\it via} roll-orienting effects which are not completely specified by ${\Gamma}$ alone.  We shall show that these effects are unimportant to SDC, provided ${\Gamma}$ is sufficiently large.   Finally, the non-Oberbeck-Boussinsesq (non-OB) parameter\cite{Bu67} ${{\cal P}}$, defined in the next section, characterizes the degree to which the fluid properties vary over the height of the layer.  Provided this parameter is small enough, as will be the case for the experimental conditions we consider here, the Boussinesq approximation\cite{Ma90} applies, and the pattern near onset consists essentially of rolls.  

SDC was first observed in gas convection using ${CO_2}$ under non-OB conditions, that is when ${{\cal P}}$ was not small.\cite{MBCA92}
In this situation, the onset of convection is a transcritical bifurcation to a pattern of hexagonal flow cells\cite{Bu67,BDAC91}.  This pattern became unstable at ${\epsilon \approx 0.1}$ to a roll pattern which, at higher ${\epsilon}$, became the disordered time dependent SDC  state.  In later experiments\cite{MBCA93}, it was found that SDC persisted under OB conditions. These conditions can be achieved experimentally at higher gas pressures\cite{MBCA93}, or in thicker samples\cite{HEA93,HEA94,LCA96}.  In the experiments with circular cells discussed here, for which ${\Gamma = 74.6}$, SDC begins to appear near ${\epsilon \approx 0.25}$ and is well developed by ${\epsilon \approx 0.5}$.  We found that this state is robust in the sense that it is found in both circular and square cells, independent of the temporal history of ${\epsilon}$, and with various lateral boundary conditions.  Similar experiments have also identified this state in circular cells with both smaller\cite{HEA93,HEA94,LCA96} and larger\cite{AS94,WAC96} aspect ratios. SDC has been observed using several other pure gases with Prandtl numbers near one, including SF$_6$ \cite{AS94,LCA96}, N$_2$\cite{PBunpub}, and Ar\cite{LCA96}. In experiments near the critical point of SF$_6$, Assenheimer and Steinberg\cite{AS94} found that SDC evolves into a state of ``target chaos" as the Prandtl number is increased.  SDC also has been studied in experiments in which there is an additional control parameter, namely the  rotation rate of the convection cell about a vertical axis\cite{EHMA95}.  It also has been observed recently in various gas {\it mixtures}\cite{LCA96} which have smaller Prandtl numbers than pure gases.  SDC has been found in several  numerical simulations of convection patterns, as modelled by modified Swift-Hohenberg equations\cite{BFFH93,XGV93,CT95}. Decker {\it et al.}\cite{DPW94} were able to quantitatively reproduce the SDC state observed in our experiment\cite{MBCA93} by a Galerkin truncation of the solutions to the Navier-Stokes equations.

The apparently ubiquitous nature of SDC may seem at first to be in conflict with the well-known picture developed by Busse and Clever\cite{CB74,BC79} of the secondary instabilities available to straight convection rolls. However, their analysis tested the stability of {\it straight roll} solutions of the convection equations to arbitrary perturbations. The SDC state, which involves defects and strong roll curvature in an essential way, appears to correspond to a different attractor that coexists with that of straight rolls over a range of parameters.  It is apparent from our experiments with square cells, discussed below, and from previous studies by Croquette\cite{Cr89}, that defect-free straight rolls represent a rather special situation that is extremely difficult to prepare in large-${\Gamma}$ cells without special forcing schemes. With small-amplitude random initial conditions, the SDC attractor is always found in simulations\cite{pesch_private}. In the experiment, random fluctuations\cite{WAC95} which nucleate the growth of convection patterns, or the roll curvature induced by sidewalls in round cells or by defects and grain boundaries in square cells, invariably seem to put the system into the attractor basin of SDC. We remark that one of the outstanding problems in the study of STC is how to generalize the concept of ``basin of attraction", which here can only be understood rather loosely in analogy with low-dimensional chaos.  A direct comparison of the SDC state with the stability boundaries given by the Busse-Clever analysis has been presented previously\cite{MBCA93,HEA95b}.

In the absence of a general framework for the understanding of STC, we fall back on statistical descriptions of the SDC state.  There is however, every reason to believe that SDC is deterministic in the sense that the external noise driving the system is quite small\cite{WAC95} and that the observed fluctuations represent the deterministic outcome of complex dynamics intrinsic to the system.   In order to characterize the statistical structure of SDC, we made use of the space-time structure of the state, in the form of time series of snapshots of the pattern.  We analyzed these using the three-dimensional structure factor ${S}({\vec k},\omega )$, which we define to be the modulus squared of the Fourier transform of the space-time shadowgraph data.  Many other sorts of analysis are possible, but this method has the virtue that the results may be interpreted in a nearly model-independent way to extract the global spatial and temporal scales of the dynamics.  This is analogous to familiar and standard techniques for the study of disorder in materials by, for example, scattering experiments.  A great deal of complexity is suppressed by this procedure, which amounts to a massive averaging over the chaotic fluctuations.  The central and open problem of STC remains to understand this complexity by a more specific reduced description.

The rest of this paper is organized as follows; in Sec. \ref{experiment} we describe the apparatus. In Sec. \ref{SDCsection} we describe in several subsections the phenomenology of spiral defect chaos, first qualitatively, and then quantitatively using the structure factor.  In these sections, we also discuss the main results on the spatial and temporal scales of the chaotic state.  Sec. \ref{discussion} presents a discussion of SDC in the context of the general features expected of STC.  Finally, Sec. \ref{conclusion} is a short conclusion.

\section{The Experimental Apparatus}
\label{experiment}

The experimental apparatus was one of several of similar design that have been used in other experiments\cite{MBCA93,HEA93,HEA95a,HEA95b,EHMA95,HEA95c,MBCA92,BDAC91,HEA94,LCA96,WAC96}.  A comprehensive description has been given\cite{DBMTHCA96}; here we describe only some essential features. The top surface of the convection cell was a circular 9.5 mm thick, optically flat sapphire window.  The window was temperature controlled by circulating water held at the same pressure as the convecting gas.  This avoided the deformation of the window that would result from a pressure differential across it, and made cells with an extremely uniform thickness possible.  The residual distortion in the flatness of the sapphire, which was ${\approx 1 \mu m}$, was due to stress from its mechanical mounting.  A pressure vessel enclosed the cell and a closed volume of water which was circulated over the top plate by a small pump. The temperature of the circulating water was fixed at $24.00 \pm 0.02 ^{\circ} C$ and regulated by a computer controlled servo.  The water pumped across the top sapphire could be regulated to $\pm{0.2}{mK}$.  

The bottom surface of the cell was an aluminum plate, 9.5 mm thick, which was diamond machined to ${\approx 1 \mu m}$ surface flatness.  The plate was heated by a commercial thin film heater which covered its entire bottom surface.  Its temperature was sensed by an embedded thermistor.  The temperature of the plate, which was under the control of a computer servo similar to that used for the circulating water, was the main experimental control parameter.  It varied between about $30^{\circ} C$ and $45^{\circ} C$, measured to $\pm 0.02 ^{\circ} C$, and was regulated to $\pm{0.2}{mK}$. The position of the bottom plate was determined by an adjustable support which included three piezoelectric legs. These allowed adjustment of the cell thickness over a range of ${\approx 10 \mu m}$, and were used as a fine adjustment after the cell was pressurized.  The height of the cell and its uniformity were measured interferometrically at the working pressure, as described in Ref. \cite{DBMTHCA96}. The cell-height uniformity was typically ${\pm 1 \mu}$m.  The cell thickness could be determined to within ${ \pm 1 \mu}$m at a reference bottom plate temperature.  As the bottom plate temperature was increased, the cell thickness changed, due primarily to the thermal expansion of the aluminum bottom plate. This expansion was about $1\%$ of $d$ at the highest bottom-plate temperatures. It was measured, and a correction was applied in the calculation of ${\epsilon}$ described below.

The lateral sidewalls of the cell were defined by a porous gasket made of three sheets of filter paper. The relatively low thermal conductivity of the paper reduced temperature gradients caused by the tendency of the gasket to act as a thermal short circuit between the top and bottom plates. The effect of such gradients on patterns in circular cells at low ${\epsilon}$ is to stabilize cell filling spirals\cite{BDAC91,PB96} or concentric target patterns\cite{KP74,SAC85,Cr89,HEA93,PB96} in which the rolls lie parallel to the sidewall.  We constructed both circular and square cells using paper sidewalls for which the  gradients were sufficiently small that the rolls lay perpendicular to the sidewalls except at very small ${\epsilon}$.  In order to study the effect of sidewall forcing,  we embedded fine heater wires in the paper sidewall which were sufficient to restore strong parallel roll alignment near onset. In the circular cell, the sidewall heater extended around the entire circumference, while in the square cell there were heaters on two opposite edges.  The circular cell was ${43.88 \pm 0.05}$ mm in radius and ${588 \pm 1 \mu}$m high, giving an aspect ratio ${\Gamma = }$ radius/height ${\sim 74.6}$.  The square cell was  ${62.5 \pm 0.1}$ mm on each side and ${639 \pm 2 \mu}$m high, so that the aspect ratio ${\Gamma = }$ width/height ${= 98}$.  In both the round and the square cells, the ${CO_2}$ gas was held at a constant pressure of $33.1 \pm 0.1$ bar. The pressure was regulated to be constant to within ${\pm 0.01\%}$ by servo controlling a heater in a gas filled ballast volume connected to the cell.  

The convection patterns were imaged using the shadowgraph technique\cite{DBMTHCA96,SAC85} which maps the deviation of a parallel beam of light reflected off the mirrored bottom plate.  The deviations are caused by the varying index of refraction of the gas due to the small lateral temperature differences across a convection roll.  In these images, the white areas correspond to cool downward flowing gas, while the dark areas are warm upflows.  To make quantitative pattern measurements, the shadowgraph was operated in a linear regime and the image was captured by a CCD video camera using a computer and an 8 bit frame grabber.  Dividing the image by a reference image taken below the onset of convection largely compensated for nonuniformities in the illumination.  Our quantitative results are based directly on these ratio images. For the purpose of visual presentation, the ratio images were re-scaled so as to render them with increased contrast.

Values of the relevant dimensionless parameters were calculated from the known gas properties.\cite{DBMTHCA96} The properties were evaluated at the average temperature ${\bar{T} = (T_t + T_b)/2}$ where $T_t$ and $T_b$ are the temperature at the top and bottom of the fluid layer respectively. The temperature $T_b$ was that measured by the thermistor embedded in the bottom plate.  The temperature ${T_t}$ was slightly larger than that of the circulating water because of the finite thermal conductivity of the sapphire top plate, which was taken into account. At the onset of convection this difference was 0.06$^\circ$C.  We calculated the vertical thermal diffusion time scale ${t_v = d^2/\kappa}$ as a function of ${\bar{T}}$ and used it to rescale all the times in the experiment.  The non-OB parameter ${\cal P}$ is given by\cite{Bu67}
\begin{equation}
{\cal P} = {\Delta T}\biggl[{p_0 \alpha} + {p_1\over{2\rho\alpha}} \biggl ({\partial^2{\rho}\over{\partial{T}^2}}\biggr ) + {p_2\over{\eta}}\biggl ({\partial{\eta}\over{\partial{T}}}\biggr ) +  {p_3\over{\Lambda}}\biggl ({\partial{\Lambda}\over{\partial{T}}}\biggr )  + {p_4\over{C_P}}\biggl ({\partial{C_P}\over{\partial{T}}}\biggr )\biggr]_{T=\bar{T}}\label{nobparam}
\end{equation}
where ${\alpha}$ is the isobaric thermal expansion coefficient $-(1/ \rho)(\partial \rho / \partial T)_P$, $\Lambda$ the thermal conductivity, and $C_P$ the heat capacity at constant pressure. The coefficients ${p_i}$ are dimensionless linear functions of ${\sigma^{-1}}$ and are given in Ref. \cite{Bu67}.  The derivatives  in Eq.\ (\ref{nobparam}) were evaluated numerically from the fits to ${\rho}$, ${\eta}$, ${\Lambda}$ and ${C_P}$ given in Ref. \cite{DBMTHCA96}.  Since only ${T_b}$ was changed while the temperature of the circulating water bath was held fixed, the mean temperature ${\bar{T}}$ changed and consequently ${\sigma}$, ${t_v}$ and ${\cal P}$ all varied slightly with ${\epsilon}$.

In the circular cell, we found the onset of convection at $\bar T_{on} = 27.32^\circ$C and ${\Delta T_c}^{expt}= 6.513 {\pm} {0.005^{\circ}}$C. The result for ${\Delta T_c}^{expt}$ agrees to within about 3\%  with the value ${{\Delta T_c}^{fit}(\bar{T}_{on})} = 6.74^\circ$C calculated from the gas properties\cite{DBMTHCA96} and the thickness.  In order to calculate ${\epsilon}$ well above onset, we must take into account the variation of the fluid parameters, and hence ${\Delta T_c}$,  with ${\bar{T}}$.  We define ${{\epsilon} = [{{\Delta T}/{\Delta T_{c0}}(\bar{T})}]-1}$, where ${\Delta T_{c0} (\bar{T})}$ is the critical temperature difference for a fluid with properties appropriate to the mean temperature ${\bar{T}}$.  To remove the effect of the small systematic difference between ${\Delta T_c}^{expt}$ and ${{\Delta T_c}^{fit}(\bar{T}_{on})}$, we defined ${\Delta T_{c0} (\bar{T})}$ as
\begin{equation}
 {\Delta T_{c0 }(\bar{T})}={\Delta T_c}^{fit}(\bar{T}) + [ {\Delta T_c}^{expt} - {\Delta T_c}^{fit}(\bar{T}_{on})].\label{dt_defn}
\end{equation}
Thus, we use fits to the gas properties to get the functional form of ${\Delta T_c}^{fit}(\bar{T})$, which is nearly linear, and include the small correction term in square brackets in Eq.\ (\ref{dt_defn}) so that this curve passes through the experimentally observed value of ${\Delta T_c}^{expt}$.  This correction becomes relatively unimportant as $\epsilon$ becomes large.
Using this definition of ${\epsilon}$, the control parameter dependence of ${\cal P}$ in the circular cell is given approximately by ${{\cal P} \approx  -0.94-1.22\epsilon}$, while the mean Prandtl number was ${\sigma = 0.94}$, and the vertical thermal diffusion time was ${t_v = 1.44s}$. For our range of $\Delta T$, ${\sigma}$ varied by 3\% and ${t_v}$ by 10\% due to the variation of ${\bar{T}}$.

\section{Spiral Defect Chaos}
\label{SDCsection}

\subsection{ Qualitative features }
\label{qualitative}

In this section, we describe the general features of the SDC state, and how they depend on some simple variations of the experimental conditions.  MBCA96Figures \ref{patternsequence_ab} and \ref{patternsequence_cd} shows the sequence of patterns observed in a circular cell with the sidewall heater turned off.  For small $\epsilon \alt 0.05$, as shown by Fig. \ref{patternsequence_ab}a, the pattern consists of motionless straight rolls, with small grain boundaries around the circumference. This is consistent with the prediction of Schl\"uter, Lortz, and Busse\cite{SLB65} and the weak forcing nature of the paper sidewalls.  As ${\epsilon}$ is increased, the tendency of the sidewalls to enforce a perpendicular roll orientation becomes more pronounced, eventually leading to the frustrated pattern shown by Fig. \ref{patternsequence_ab}b at ${\epsilon}{~~}{\simeq}{~~}{0.245}$.  These patterns typically show at least {\it three} focus singularities around their circumference and several slow-moving defects and grain boundaries in the interior. The foci often act as sources or sinks for rolls and dislocations\cite{HEA94}.  This state may be contrasted with the two-focus ``Pan-Am" pattern seen in smaller $\Gamma$ cells\cite{SAC85,Cr89,HEA93} in this regime.  At ${\epsilon}{~~}{\approx}{~~}{0.25}$, the first spiral defects are seen. These appear quite intermittently, and usually develop within highly defected regions near the center of the cell.  The spirals exhibit a tendency to rotate and evolve on shorter time scales than the rest of the pattern.  By ${\epsilon}{~~}{\approx}{~~}{0.5}$, the nucleation of spiral defects is continuous, usually in the interior of the cell, and the highly defected SDC regions coexist with slow-moving focus singularities around the edge of the cell, as shown by the Fig. \ref{patternsequence_cd}a.  As ${\epsilon}$ is increased, the SDC state progressively invades the entire cell, as shown by Fig. \ref{patternsequence_cd}b.  Even at the highest ${\epsilon}{~~}{\approx}{~~}{1.4}$,  the rolls continue to meet the sidewall perpendicularly, creating a circumferential fringe of small focus defects as long as the sidewall heater is turned off.

We may note several qualitative features of the SDC state, as it appears in the interior of the cell, away from boundary effects.  First, we find that both right and left handed spirals appear, in roughly equal numbers, and are {\it not} created in pairs. In fact, the spiral cores are only the most visually striking defect in a pattern which contains many other defects including dislocations, grain boundaries {\it etc.} One-armed spirals are certainly the most common, but target patterns and two-armed spirals also appear occasionally.  The processes by which spirals are created and destroyed are not simple in general, but some common scenarios have been described by Assenheimer and Steinberg\cite{AS94}.  The rotation rate of the spirals is variable and difficult to estimate except in a few particularly long-lived examples.  In cases where a sustained rotation is seen, the spiral core rotates in a ``winding-up" sense, {\it i.e.} so that with successive turns rolls move outward from the core.  

The bottom plate temperature could either be ramped slowly, or stepped rapidly.  We are only concerned here with patterns which have reached a statistical steady state.  In the absence of other information, a pattern transient may be expected to take at least several horizontal thermal diffusion times ${t_h \equiv \Gamma^2 t_v \approx 2.2h }$. In fact, over most of the range of ${\epsilon}$ we studied, the SDC state takes much less time than this to establish itself, requiring only a few hundred ${t_v}$.  This is apparently due to the surprisingly fast time scale of SDC, as we will make quantitative below.  Nevertheless, our experimental protocol was to step ${\epsilon}$ rapidly (within ${\approx 10 t_v}$) to a target value, then to wait at least ${2 t_h}$ before collecting data.  Runs in which ${\epsilon}$ was ramped slowly, $({t_v} {{d}{\epsilon}\over{dt}} {~~}{\simeq} {~~}10^{-5})$, produced similar patterns at low ${\epsilon}$ and a similar transition to SDC.

In order to examine the effect of changing the sidewall boundary conditions, we performed a few runs with the sidewall heater on. For small ${\epsilon} \alt {0.20}$, the main effect is to convert straight roll and frustrated patterns similar to those of Fig. \ref{patternsequence_ab} to large cell-filling spirals, as shown in Fig. \ref{roundcellheaterloweps}.  These large spirals were qualitatively similar to the rotating spirals observed previously\cite{BDAC91,PB96}.  If the sidewall heater is turned off after a cell-filling spiral is established, it was found that the center of the spiral soon moves to one side and a frustrated pattern similar to that shown in the Fig. \ref{patternsequence_ab}b develops after a long transient.  At higher ${\epsilon}$, with the sidewall heater on, a similar SDC state is found, as shown in Fig. \ref{roundcellheaterchaos}. Here SDC evolves in the interior of the cell, while at the perimeter the rolls lie parallel to the sidewalls and no small foci are present.  We have not yet undertaken a detailed study of the transition between the large spiral states\cite{BDAC91,PB96} and SDC. It would be interesting to compare the onset of SDC in a circular cell with and without sidewall forcing, and in various aspect ratios, using the methods of Ref. \cite{HEA94}.  

To test the sensitivity of the SDC to the sidewall geometry, we examined the phenomenon in our square cell.  Figure \ref{sqcell} shows the patterns observed. We attempted to use the sidewall heaters on opposite edges to prepare a state of straight rolls, but this proved to be impossible even for very slow ramping of ${\epsilon}$ through onset.  This was apparently due to a very narrow range of hexagonal flow cells that appears just above onset, due to non-OB effects; these tend to decay to roll patches that meet at ${120^{\circ}}$, leading to persistent grain boundaries as in Fig \ref{sqcell}(a).  We also tried deliberately breaking the symmetry by putting a large thickness (and hence ${\epsilon}$) wedge across the cell from edge to edge, and by tilting the entire apparatus several degrees toward one edge.  None of these methods were sufficient to prepare straight rolls.  While this effort would probably be successful with a sufficiently drastic forcing scheme,  it seems fair to say that defect-free straight rolls are practically inaccessible given the initial conditions and boundary conditions that were experimentally available in our cell. They have been successfully prepared in smaller aspect ratio rectangular cells using sidewall heaters, however\cite{Cr89}. At higher ${\epsilon~\simeq~ 0.5}$, we found the familiar SDC state in the square cell, regardless of the sidewall heating, as shown in Fig. \ref{sqcell}(b).

The observation of SDC with and without sidewall forcing in circular cells, in cells of various ${\Gamma},$\cite{HEA93} in sufficiently large square cells, and in Navier-Stokes simulations with periodic boundary conditions\cite{DPW94} lends support to the conclusion that SDC is a generic state for low-${\sigma}$ and large-${\Gamma}$ convection, which is not related to the detailed boundary conditions. In the next section, we take up the quantitative characterization of SDC.  This will show that the characteristic length scales in the SDC regime are much smaller than the size of the cell, as one expects for an ``extensive" chaotic state.

\subsection{ Quantitative analysis using the structure factor.}
\label{quantitative}

In this section we describe the statistical methodology we used to quantitatively characterize the SDC.  A time series of two-dimensional shadowgraph snapshots is taken and may be regarded as a sample of three dimensional space-time. We presume that the states for which we take such series are statistically stationary.
Our basic tool is the structure factor ${S}({\vec k}, \omega )$ which we determine experimentally by taking the modulus squared of the discrete Fourier transform of the three-dimensional space-time data set.  Since phase information is destroyed by taking the modulus, this already represents an enormous simplification of the data.  We further simplify the structure factor using certain windowing and averaging techniques described below, to extract a few global features of the motion, namely correlation lengths and times, as a function of ${\epsilon}$.  This analysis leaves much to be desired and averages out a lot of the interesting details of the dynamics; it has the virtue of simplicity and is almost completely free of {\it ad hoc} assumptions.

Two types of spacetime data were collected.  In the first type, time series were taken with a long wait, typically several hundred ${t_v}$, between images.  This time step was chosen to be long compared to the typical timescale for pattern change, so that successive images were nearly uncorrelated. This type of data was used for obtaining long time averages of ${S}({\vec k}, t )$.  A second type of time series consisted of images taken a few ${t_v}$ apart, so as to obtain temporal correlation information.  In both cases, each divided image was ${256{~}\times{~}256}$ pixels with 8-bit greyscale resolution. One spacetime sample usually contained 256 such images.  The size of the square image was such that its corners just met the edges of the round cell. In order to avoid including the pattern in the corners of the image, which is influenced by boundary effects, and to reduce the aliasing effect of the sharp edged picture, we prefiltered the images by multiplying them by a radial Hanning function given by 

\begin{equation}
H(r){~~}{\equiv}{~~}{[{1}+ {cos}({\pi} { r} / {r_0})]/2},{~~~~~~}{r \leq {r_0}}\label{hanninginside}
\end{equation}
\begin{equation}
{H(r) }{~~}{\equiv}{~~}0,{~~~~~~}{r > {r_0}}.\label{hanningoutside}
\end{equation}

The radius ${r_0}$ was ${0.71 {\Gamma}}$ in units of $d$.  A one-dimensional Hanning window was also applied to the spacetime data in the time direction.  The effect of each of these windows is to convolve the windowing function with the Fourier transform in such a way that ringing due to the sharp boundaries of the dataset is suppressed.

\subsection{The time averaged structure factor }
\label{timeavg}  

We consider the {\it time averaged} structure factor ${S}({\vec k})$ for the temporally uncorrelated spacetime data first\cite{MBCA93}.  This could either be obtained by numerically integrating ${S}({\vec k},\omega )$ over ${\omega}$, or more simply by time averaging the instantaneous spatial structure factors ${S}({\vec k}, t)$at each time step.  The latter is equivalent to the former.  When SDC is well developed, ${S}({\vec k})$ shows a broad ring centered on ${\vec k = 0}$.  Only data within an annular band near this ring were further analyzed; we ignored, for example, a large spike near ${\vec k = 0}$,  which is due mostly to large scale nonuniformities of illumination.  A diffuse positive background, presumably due to camera noise, pervades ${\vec k}$-space, and underlies the ring.  The shadowgraph was operated in as nearly linear a regime as possible, but at the highest ${\epsilon}$, a trace of a second harmonic ring was detectable.  This ring was either due to residual nonlinearity in the shadowgraph, or possibly higher order components within each roll of the flow pattern itself.  In any case, only data near the primary ring were considered, and the very small tail due to the second harmonic was treated as part of the diffuse background. 

In the vicinity of the onset of SDC, for ${\epsilon} \alt 0.3$, the ring in ${S}({\vec k})$ appears irregular or broken, while near the onset of convection it is reduced to two spots.  For ${\epsilon} \agt 0.4$, where spirals appear continuously, ${S}({\vec k})$ was nearly azimuthally symmetric.  Hu {\it et al.}\cite{HEA94} have shown that the uniformity of the power around the ring is a useful diagnostic for the appearance of spirals and hence of SDC. This is obviously due to the fact that spirals or targets naturally contain rolls of all orientations, while the frustrated patterns similar that in Fig. \ref{patternsequence_ab}b usually contain a dominant roll orientation.  More precisely, nonuniform ${\vec k}$-space rings result when the slow roll rearrangements in the frustrated patterns have insufficient time to sample all roll orientations within the averaging time of the dataset.  Once the quickly-evolving spirals begin to reliably appear, the ring fills in over relatively short averaging times.

The inverse width of the ring is a quantitative measure of the length scale of the patches of rolls in the pattern.  We performed an azimuthal average in ${\vec k}$-space to obtain the azimuthally and time averaged structure factor ${S}({k})$, where ${k}{~~}{\equiv}{~~}|{\vec k}|$. Using the average wavenumber $\langle k \rangle$ from this distribution, defined by
\begin{equation}
\langle k \rangle{~~}{\equiv}{~~}{{{\int}|{\vec k}|{S}({\vec k}) {d^2}{\vec k}}\over
{{\int}{S}({\vec k}){d^2}{\vec k}}} = {{{\int_0^{\infty}}{k^2}{S}({k}){dk}}\over
{{\int_0^{\infty}}{k} {S}({k}) {dk}}}\ ,\label{skmean}
\end{equation}
we define the moments $\mu_n$ of ${S}$ about $\langle k \rangle$ as
\begin{equation}
\mu_n{~~}{\equiv}{~~}\Biggl[{{{\int}
(|{\vec k}|-\langle k \rangle)^n 
{S}({\vec k}) {d^2}{\vec k}}\over
{{\int}{S}({\vec k}){d^2}{\vec k}}}\Biggr] =
\Biggl[{{{\int_0^{\infty}}
( k-\langle k \rangle)^n
k{S}({k}){dk}}\over
{{\int_0^{\infty}}{k} {S}({k}) {dk}}}\Biggr]\ .\label{moments}
\end{equation}
The correlation length ${\xi}$ is calculated from the second moment via 
\begin{equation}
\xi=1 / \sqrt{\mu_2}\ .
\label{corrlength}
\end{equation}
 
The values of higher moments extracted from data are very sensitive to noise in the wings of ${S}({k})$.  Reliable values up to the fourth moment were extracted from ${S}({k})$ by first fitting $k S(k)$ to a smooth phenomenological model function, given by
\begin{equation}
kS(k) = [a_0 + a_1 k] + [a_2 + a_3 k] e^{-[a_5(k-a_4)^2 + a_6(k-a_4)^3 + a_7(k-a_4)^4]}\ ,
\label{modelfun}
\end{equation}
using a nonlinear least squares algorithm, where the $a_i$ are adjustable parameters.  The $a_0$ and $a_1$ terms of the model function are a linear background contribution, while the rest form a skewed gaussian ``bump" piece which decreased exponentially in its wings.  With eight parameters, an essentially perfect fit could always be achieved.  Once the data were fit, moments were extracted numerically from {\it the model function}, using only the second term in Eq. \ref{modelfun}. This method was robust to noise in the wings, modifications of the fitting range, and omissions of points in the data set. It removed the background contribution as well.  

\begin{figure}
\epsfxsize = 5in
\centerline{\epsffile{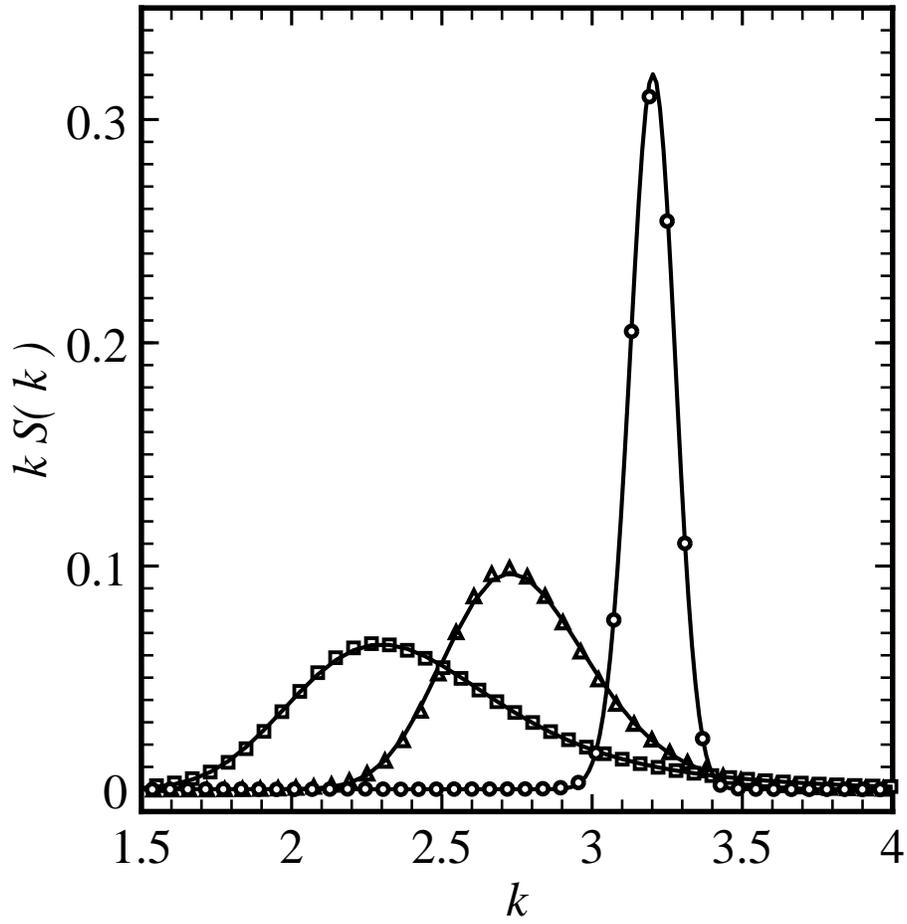}}
\vskip 0.5in
\caption{The time average of $kS(k)$ for $\epsilon=0.043$ (round points), $\epsilon=0.306$ (triangles), and $\epsilon=1.032$ (squares). }
\label{tavgSk3eps}
\end{figure}

Figure \ref{tavgSk3eps} shows the structure factor that results from this averaging procedure for several values of $\epsilon$. The solid lines are the fits of Eq. \ref{modelfun} to the data. For the lowest $\epsilon$ shown, the rolls are essentially perfectly straight over the whole region sampled.  Nevertheless, ${S}({k})$ still has some width due to the finite number of rolls in the sample, and the effect of the Hanning window.  This may be regarded as a ``resolution-limited" structure factor; it corresponds to our maximum resolvable correlation length $\xi \approx 14 d$. As $\epsilon$ is increased, $\langle k \rangle$ shifts to smaller $k$, and $S(k)$ becomes significantly skewed to higher ${k}$ in the SDC regime, as we discuss below.

In a previous publication\cite{MBCA93}, we compared the mean wavevector $\langle k \rangle$ of the pattern to the band of stable wavevectors expected for straight rolls, according to Busse and Clever\cite{CB74,BC79}. The trend in $\langle k \rangle$ shows no strong changes near the onset of SDC, and roughly tracks the middle of the predicted stable band, starting from onset.  This may be contrasted with the results of Hu {\it et al}, who found significant deviations from the middle of the band in a $\Gamma=40$ cell.\cite{HEA95a,HEA95b}  Evidently, the processes that select $\langle k \rangle$ at various $\epsilon$ depend on $\Gamma$.

\begin{figure}
\epsfxsize = 5in
\centerline{\epsffile{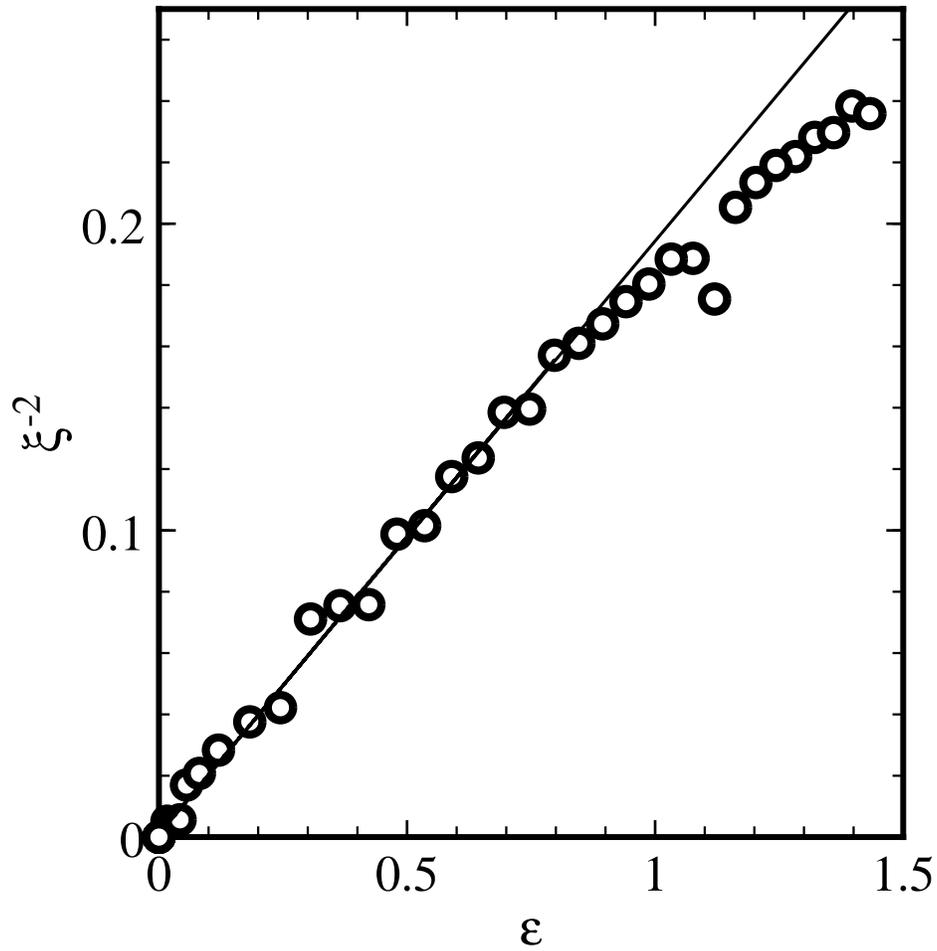}}
\vskip 0.5in
\caption{A plot of $1/\xi^2$ {\it vs.} $\epsilon$. The solid line corresponds to $\xi^{-2} = 0.194 \epsilon$. The good fit to the data demonstrates that $\xi \sim \epsilon^{-1/2}$ for small $\epsilon$. }
\label{corrlensquared}
\end{figure}

The correlation length ${\xi}$ decreases with increasing ${\epsilon}$\cite{MBCA93}, and our results can be fit by the power law 
${\xi} \propto {\epsilon^{- 0.43 \pm 0.05}}$. To show that the exponent of this power law is consistent with $1/2$, we plot ${\xi^{-2}}$ {\it vs.} ${\epsilon}$ in Fig. \ref{corrlensquared}. If ${\xi}$ is exactly proportional to ${\epsilon^{-{1 / 2}}}$, this will be a straight line passing through zero.  As Fig. \ref{corrlensquared} shows, this is quite consistent with the data, with  small systematic deviations only at larger ${\epsilon}$. A fit of $\xi ^{-2} = \xi_0^{-2} \epsilon$ to the data at small $\epsilon$ is shown by the solid line in Fig. \ref{corrlensquared}, and gives $\xi_0 \simeq 2.3$ (in units of $d$).  We have at present no theoretical insight into why this power law should appear, why the exponent should be so close to $1/2$, and why $\xi$ should diverge at $\epsilon = 0$ rather than, say, at some small positive $\epsilon$.  The correlation length is presumably unrelated to the one which appears in an amplitude equation valid near $\epsilon = 0$. That correlation length has the same exponent 1/2, but has $\xi_0 = 0.385$ which is much smaller than our result.

The qualitative meaning of the correlation length as it relates to the interpretation of images requires some clarification\cite{Hu95}.  It is tempting to assume that ${\xi}$ is a measure of the typical spiral diameter, but this is incorrect.  A disordered pattern consisting of a patchwork of nearly ordered subpatterns of typical size ${\ell}$ will only exhibit a correlation length ${\xi \approx \ell}$ if the patches consist of nearly {\it straight rolls}.  If the patches are targets or spirals, ${\xi}$, as we have defined it, will be systematically smaller than ${\ell}$\cite{Hu95}. This is a consequence of basing the definition of ${\xi}$ on the choice of the Fourier transform ({\it i.e.} on the standard definition of the structure factor), a transform based on a kernel consisting of plane waves.  If the subpatterns have broadband structure in their Fourier transforms, this contributes to a broadening of the overall structure factor and a reduction of ${\xi}$.  The choice of the Fourier transform, which is sanctioned by long usage in condensed matter physics, is a natural one for the interpretation of scattering experiments, but does not correspond very well to what the eye naturally picks out as ``coherent structures" in SDC patterns.  Obviously, many other sorts of transform could be applied.  Also, while the eye is drawn to the spirals, it is well to remember that SDC patterns contain significant areas which are dominated by other types of defects, such as grain boundaries, dislocations {\it etc.}  The measured correlation length must be interpreted to be an average over of all of these, weighted according to their individual effect on the width of the structure factor.  Similar considerations apply to the higher moments, to which we now turn. 

\begin{figure}
\epsfxsize = 4in
\centerline{\epsffile{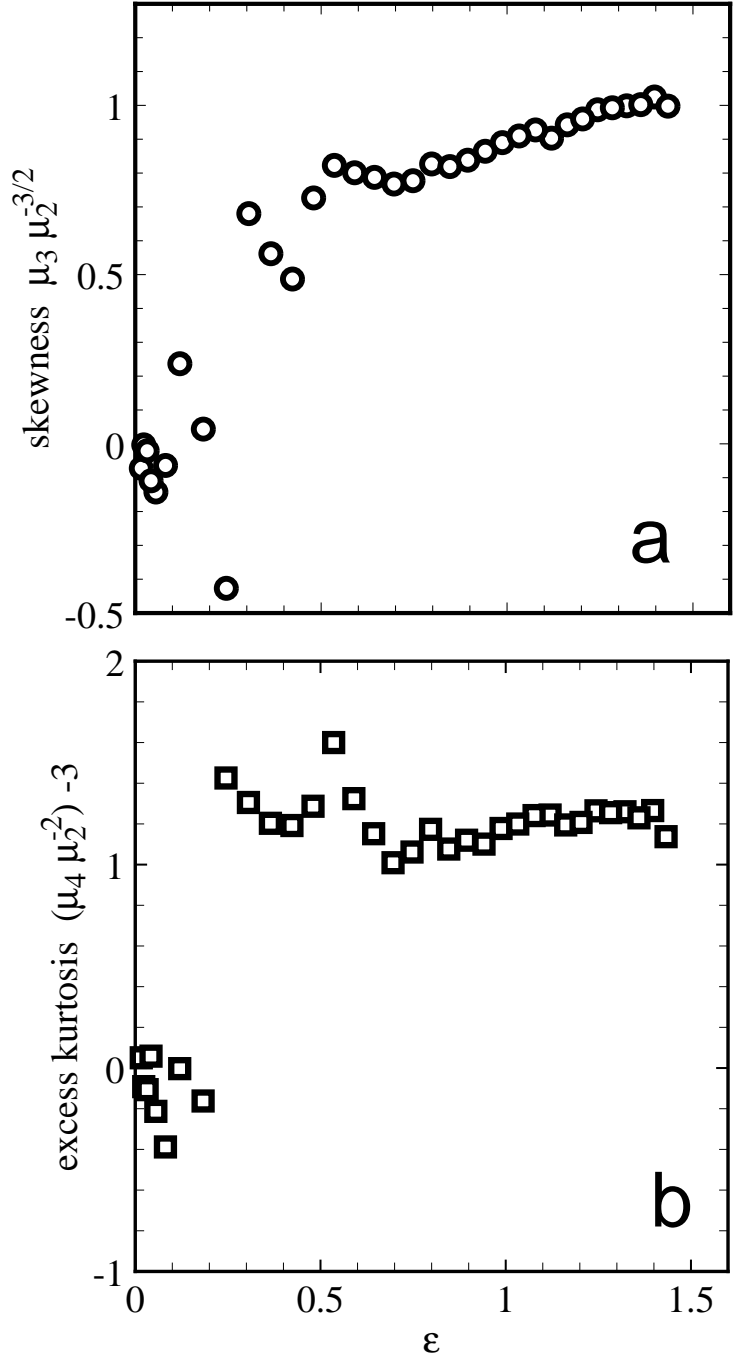}}
\vskip 0.5in
\caption{The higher moments of the time averaged structure factor, (a) the skewness and (b) the excess kurtosis, as a function of $\epsilon$.}
\label{skewkurt}
\end{figure}

From the higher moments of ${S}({k})$, we calculated the skewness $s = \mu_3 \mu_2 ^{-3/2} $ and the excess kurtosis $\kappa = \mu_4 \mu_2 ^{-2} - 3$.  The latter is defined in such a way that $\kappa$ is exactly zero for a gaussian distribution. Figure \ref{skewkurt} shows the dependence of these statistics on $\epsilon$.  For low $\epsilon$, the distribution is symmetric and indistinguishable from a gaussian, so that both $s$ and $\kappa$ are nearly zero.  The appearance of spirals near $\epsilon \approx 0.25$ is accompanied by a sudden increase in both statistics.  After a considerable transient, the skewness continues to increase as the SDC develops. The excess kurtosis shows a very sharp jump at the onset of SDC, after which it is remarkably constant.  These are nothing more than descriptive statistics, but they do contain the essential global features of the time-averaged spatial correlations of SDC that can be extracted in a model-independent way.  The skewness has previously been used as a diagnostic of the onset of SDC\cite{HEA95a,HEA95b}.  Our data indicates that the excess kurtosis gives an even clearer indication.  Qualitatively, the increases at the threshold of SDC can be understood from the considerations of the previous paragraph.  Spirals represent localized structures whose Fourier transforms contain significant asymmetric contributions in wings distributed about the local mean $k$. As the spirals appear and proliferate, $S(k)$ develops both skewness and kurtosis. The larger  scatter in the skewness for $0.2 \leq \epsilon \leq 0.5$ relative to that of the kurtosis apparently reflects the fact that the fourth moment is less sensitive to sampling statistics in the calculation of $S(k)$.  In the lower part of this $\epsilon$ range, the spirals appear only rather intermittently and the statistics of their contribution to the skewness is not very good. The extraction of moments is simply an objective way of disentangling the changes in the shape of $S(k)$. The moments, particularly the higher ones, lack any simple, direct physical interpretation. However, such statistics may be useful for quantitatively comparing our results to numerical simulations\cite{DPW94}.

\subsection{Lifetimes from the three-dimensional structure factor }
\label{lifetimes}  

We now return to the analysis of the structure factor ${S}({\vec k}, \omega )$, extracted from spacetime data sampled fast enough that successive pictures are correlated.  Some insight into the degree of temporal correlation can be obtained by plotting two-dimensional slices of spacetime, as in Fig. \ref{spacetime}.  The temporal behaviour of the pattern as it passes a particular line in space is shown spread out along the time axis.  The regions of narrow stripes correspond to rolls which happen to meet the chosen line at nearly right angles, while the broad patches are rolls nearly parallel to the chosen line. A stack of 256 pictures like Fig. \ref{spacetime} make up the three dimensional spacetime data set at each $\epsilon$. Each data set represents $2^{32}$ bits of information. The calculation of ${S}({\vec k}, \omega)$, which destroys the phase information, reduces the number of bits by a factor of two.  In addition, we ignored regions of ${\vec k-\omega}$ space far from the ring, which reduced the amount of data substantially.

\begin{figure}
\epsfxsize = 5in
\centerline{\epsffile{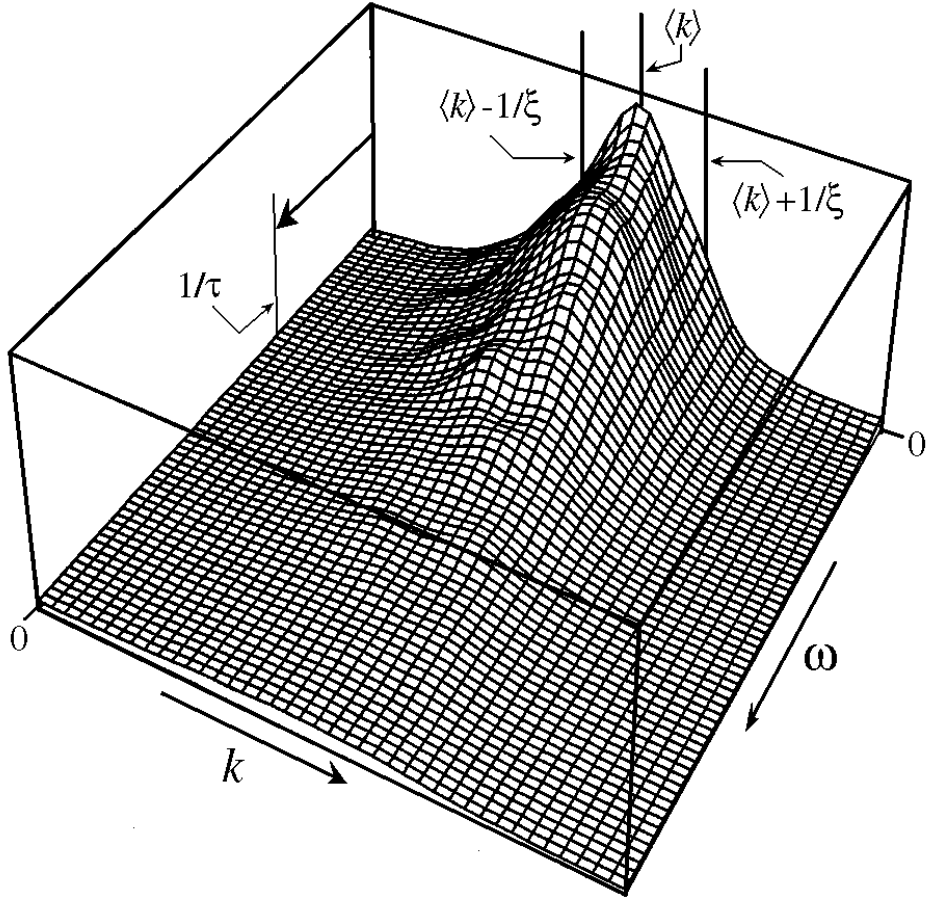}}
\vskip 0.5in
\caption{A plot of the surface of a typical azimuthally averaged $S(|{\vec {\bf k}}|,\omega)$, showing the relation between the various statistical quantities. }
\label{3dSkw}
\end{figure}

To reduce the data still further, we make use of the rotational symmetry of ${S}({\vec k}, \omega)$ about the ${\vec k = 0}$ axis, and take an azimuthal average to find $S(k,\omega)$.  An example of the resulting two-dimensional data is shown in Fig. \ref{3dSkw}.  In the ${k-\omega}$ plane, $S$ decreases monotonically in the direction of increasing ${\omega}$ for all $k$.  Because of the relatively rapid sampling, we cannot analyze as large a volume of spacetime as was the case for the data discussed in the previous section.  As a consequence, statistical fluctuations are larger.  

\begin{figure}
\epsfxsize = 5in
\centerline{\epsffile{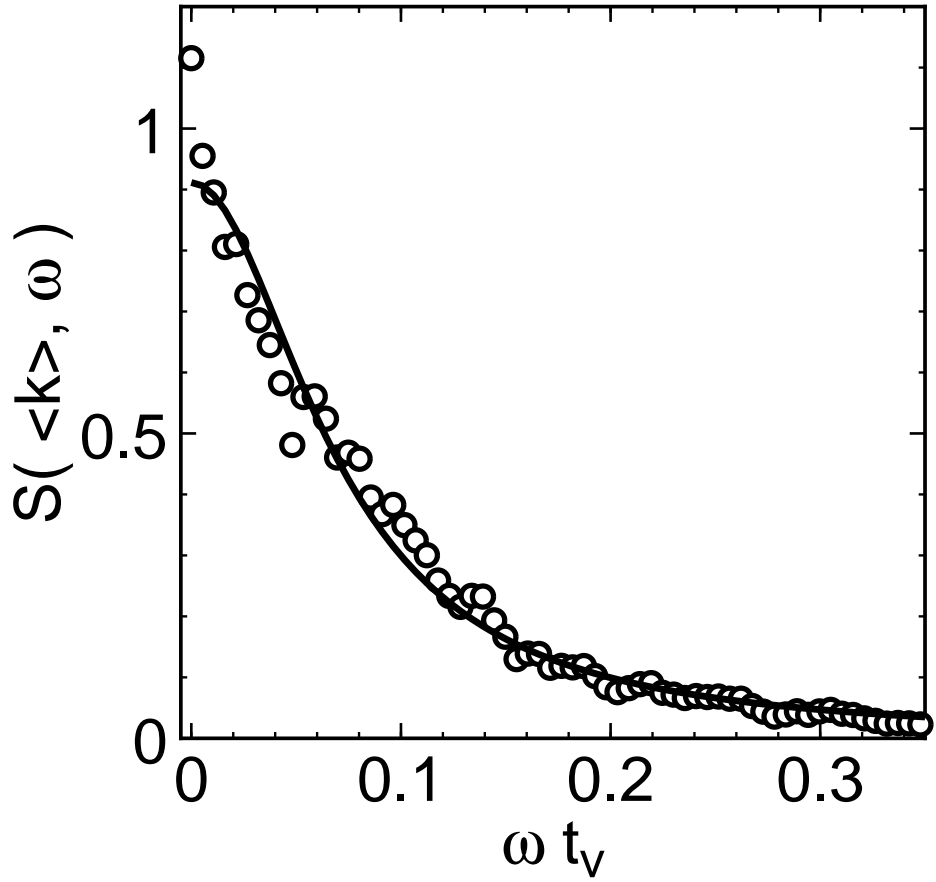}}
\vskip 0.5in
\caption{A dimensionless plot of $S(k,\omega)$ vs $\omega t_v$, for $k= \langle k \rangle$, at $\epsilon=0.747$.  The solid line is a fit to a two parameter Lorentzian. The fit gives $\tau = 12.3 t_v$.}
\label{Skwvsw}
\end{figure}

The rate of decrease of ${S}(k,\omega )$ as a function of ${\omega}$ is a quantitative measure of the lifetime, or correlation time, of structures in the pattern with characteristic size ${2\pi/k}$.  Unfortunately, sampling statistics limit the useful part of the data to $k$ values near $\langle k \rangle$, {\it i.e.} near the peak of ${S}(k,\omega )$.  Figure \ref{Skwvsw} shows a plot of ${S}(\langle k \rangle,\omega )$ {\it vs.} ${\omega}$ for $\epsilon = 0.75$.  If correlations decrease exponentially in time, this relation will be Lorentzian.  The solid line shows a fit to 
\begin{equation}
S = A/(1 + \omega^2 {\tau}^2) ,
\label{lorentzian}
\end{equation}
with the correlation time ${\tau}$ and arbitrary amplitude $A$ as fit parameters.  In many cases, however, an exponential provides a better fit, and this  also yields an arbitrary amplitude and a time scale as parameters. The times extracted from exponential fits were proportional to the correlation times found from Lorentzian fits. We adopted Lorentzian fits since they had the simplest interpretation in the time domain.   In order to improve statistics, we averaged the lifetimes found for a range of $k$ centered on $\langle k \rangle$; ${\tau}$ is only a rather weak function of $k$. 

\begin{figure}
\epsfxsize = 5in
\centerline{\epsffile{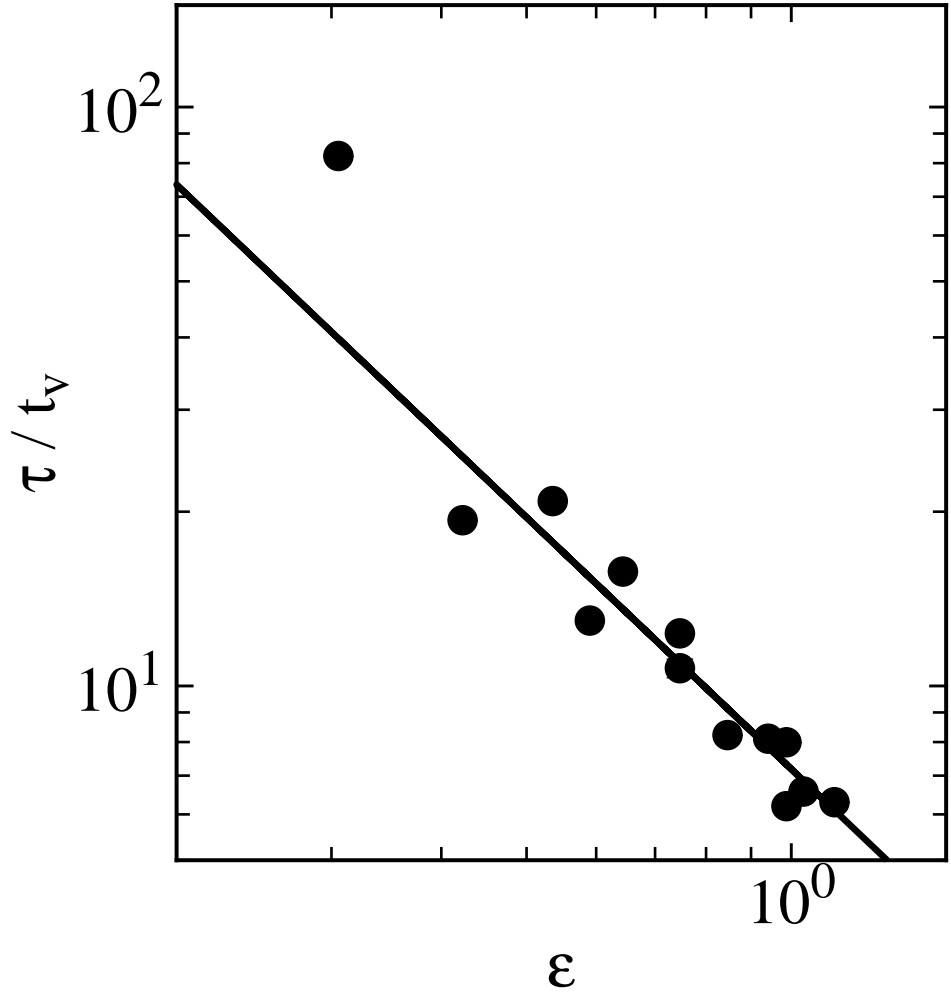}}
\vskip 0.5in
\caption{The correlation time ${\tau}$ extracted from Lorentzian fits like the one shown in Fig. {\protect \ref{Skwvsw}}, {\it vs.} $\epsilon$, for $k$ in a narrow range around $\langle k \rangle$.  The solid line is a fit which gives an exponent of ${-1.43 \pm 0.05}$.  }
\label{tauloglog}
\end{figure}

The dependence of $\tau$ on $\epsilon$ is shown on logarithmic scales in Fig. \ref{tauloglog}.  As with $\xi(\epsilon)$, we find a very rapid decrease of ${\tau}$ with  increasing ${\epsilon}$.  A fit to a power law is shown by the straight line. It gives an exponent of ${1.43 \pm 0.05}$. Thus the data are consistent with a threshold at $\epsilon = 0$ and an exponent of about 3/2. 

\begin{figure}
\epsfxsize = 5in
\centerline{\epsffile{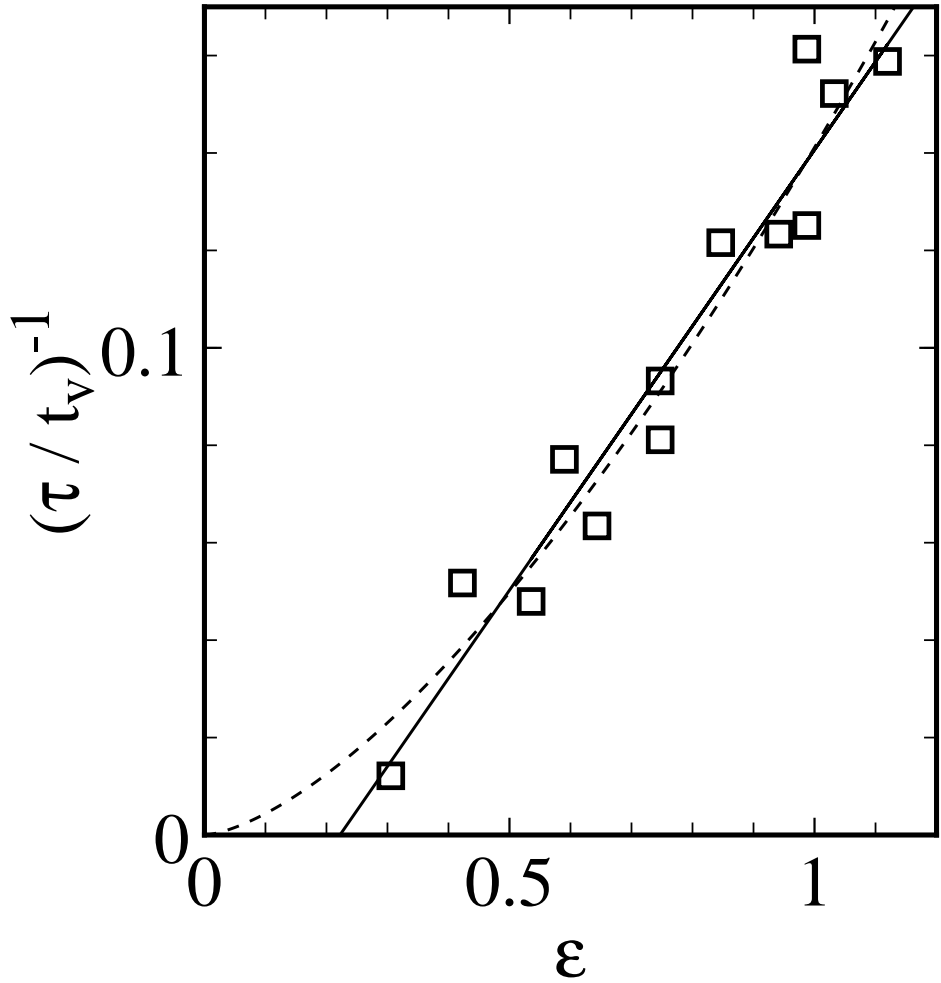}}
\vskip 0.5in
\caption{A plot of ${\tau}^{-1}$ {\it vs.} $\epsilon$, for the same data as in Fig. {\protect \ref{tauloglog}}.  This shows that ${\tau} \sim (\epsilon - \epsilon _c) ^{-1}$ with $\epsilon _c \simeq 0.22$ is consistent with the data. The powerlaw fit with an exponent of 3/2 and a threshold at $\epsilon = 0$  is given by the dashed line.}
\label{tauvsepsinv}
\end{figure}

The powerlaw analysis given above suggests a time scale which diverges at $\epsilon = 0$ as 1/$\epsilon^{3/2}$, whereas critical slowing down near bifurcation points usually involves a time scale proportional to 1/$\epsilon$. In order to see whether a 1/$\epsilon$ dependence of $\tau$ might also be consistent with the data, we show $(\tau/t_v)^{-1}$ vs. $\epsilon$ in Fig. \ref{tauvsepsinv}. Also shown, as a dashed line, is the powerlaw fit with the threshold at $\epsilon = 0$. One sees that the data can be described equally well by a straight line in Fig. \ref{tauvsepsinv}, but this line passes through a {\it shifted} threshold at a positive $\epsilon_c \simeq 0.22$. This value of $\epsilon _c$ is close to the onset of SDC given by other diagnostics as shown in Fig. \ref{skewkurt}. This interpretation of the data for $\tau$ suggests that SDC appears {\it via} a {\it bifurcation} beyond which the abundance of spiral excitations grows algebraically. Other models which have been suggested\cite{EHMA95} imply an exponential growth of the frequency of spiral occurrence. At this time we know of no theoretical explanation of the powerlaw behaviour with a threshold either at $\epsilon = 0$ or at $\epsilon _c$. It is worth noting that the data for $\xi$ shown in Fig. \ref{corrlensquared} can not be fit by a function with a non-zero threshold. 

\section{discussion}
\label{discussion}

The best established general conjecture about spatio-temporal chaos concerns the ``extensive" nature of the fractal dimension of the attractor which governs the spatio-temporal dynamics.\cite{CH92,ruelle} The dimension $D$ is conjectured to be related to the characteristic size of the system $L$ as
\begin{equation}
D = (L/\xi_{c})^d
\label{Dscaling}
\end{equation}
for ``large" systems with ${L \gg \xi_{c}}$, where ${\xi_{c}}$ is defined to be the ``chaos" correlation length. Here $d$, the spatial dimension of the system, is 2 for our convection patterns. Indeed, the validity of Eq. \ref{Dscaling} might be considered as a defining feature of STC.  Parts of a large system are presumed to be nearly dynamically independent if they are separated by a distance of order ${\xi_{c}}$.  The dimension $D$ is extensive in the sense that it scales with the volume of the system, ${D \propto L^d}$.  Each volume ${{\xi_{c}}^d}$ acts like a dynamical subsystem (whose own dimension need not be small) which is weakly coupled to the whole, and contributes additively to the overall dimension $D$.  This scaling scenario has been confirmed in a number of numerical simulations\cite{CH92,egolf,egolfnature,bohr,ohern} in which $D$ can be calculated as a function of $L$ directly for certain model dynamical equations. One can then define an intensive dimension density ${\delta = D/L^d}$ so that ${\delta = 1/{\xi_{c}}^d}$.

The important question for the analysis of experiments is the relation, if any, between ${\xi_{c}}$, the range over which the dynamics is {\it physically} correlated, and the observable {\it pattern} correlation length ${\xi}$ as defined by Eq. \ref{corrlength}.  If a simple proportionality (with a coefficient of order one) exists, we can estimate the rather high dimension of the attractor for SDC in a large but finite cell, a task which is essentially impossible by the usual time series methods used for low-dimensional systems.  Encouragingly, several large scale simulations\cite{egolf,egolfnature,bohr,ohern} have found simple relations between ${\xi_{c}}$ and ${\xi}$.  Simulations of the complex Ginzburg-Landau (CGL) equation in chaotic regimes\cite{egolf,egolfnature,bohr} have found that the correlation length of the field amplitude is simply proportional to ${\xi_{c}}$.  Within the parameter space of the CGL equation, however, one has transitions between phase- and defect-dominated chaotic states, and the overall phenomenology is not simple\cite{egolf,shraiman}.  In general, one must choose carefully among the various possible definitions of ${\xi}$ for the proportionality of ${\xi_{c}}$ and ${\xi}$ to hold.  For example, in simulations of the CGL equation\cite{egolf,egolfnature,bohr}, it was found that neither the correlation length derived from the complex field itself, nor that of its phase alone scaled linearly with ${\xi_{c}}$; only the correlation length derived from its amplitude did. A similar proportionality between ${\xi_{c}}$ and a suitably defined ${\xi}$ has also been found in a completely different coupled-map lattice system\cite{ohern}.  Again, in this system other correlation lengths were found that did not scale linearly with ${\xi_{c}}$.  There seems to be no prescription, other than a direct numerical test, for deciding which correlations lengths scale and which do not.

Under the optimistic assumption that ${\xi \approx \xi_{c}}$, Eq. \ref{Dscaling} with $L = \Gamma$ and our result that ${\xi \sim \epsilon^{-{1 / 2}}}$, imply that the dimension $D$ for SDC increases linearly with ${\epsilon}$, reaching $1400$ at ${\epsilon = 1.4}$.  That is, at ${\epsilon = 1.4}$, $1400$ correlation areas ${\pi \xi^2}$ fit in the area of the cell ${\pi \Gamma^2}$. The dimension density $\delta$, which would also be linear in ${\epsilon}$, reaches ${ \approx 0.25}$. This is comparable to that found in simulations of various model equations\cite{egolf,egolfnature,bohr,ohern}.  Of course, this analysis is predicated on the untested assumption that ${\xi \approx \xi_{c}}$, which at present could only be established by prodigious amounts of numerical effort.  In principle, the relation could be tested for SDC by finding the Lyapounov dimension $D$ for realistic Navier-Stokes simulations\cite{DPW94}, or, less ambitiously, for a modified Swift-Hohenberg model\cite{BFFH93,XGV93,CT95}, using the direct methods of Ref. \cite{egolfnature}.  Such a calculation, although very computationally expensive, is probably feasible\cite{egolfprivate}.

The correlation lengths and times of SDC at large $\epsilon$ must eventually be bounded by the cutoffs imposed by the basic scales of the convection process itself, the cell height $d$ and the vertical time scale $t_v$.  At $\epsilon \approx 1.4$, we found $\xi \approx 2d$ and ${\tau \approx 6 t_v }$.  Thus, the power law dependence on $\epsilon$ cannot persist for much larger $\epsilon$.  On the other hand, for small $\epsilon$, the experimental pattern made a crossover between SDC and the frustrated state, which is dominated by sidewall effects.  A detailed study of the $\Gamma$ dependence\cite{HEA95a} of the onset of SDC would be most interesting. A very large $\Gamma$ might be achieved by working close to the critical point of the gas\cite{AB78,AS93,AS94}, but ultimately the experimental limits would be set by practical constraints on the acquisition and analysis of the very large amount of data that would be required. 

\section{conclusion}
\label{conclusion} 

In this paper we used a simple structure-factor analysis to characterize the space-time behaviour of shadowgraph images of SDC.  The spatial and temporal correlations were obtained from the first few moments of an appropriately averaged structure factor.  The length and time scales were found to scale as  power laws in the reduced control parameter $\epsilon$. At present these power laws have no theoretical explanation. It is conjectured that the pattern correlation-length is a measure of the size of physically correlated patches in the underlying chaotic dynamics.  If this is so, the extensive dimension and the intensive dimension density of the dynamics both scale linearly with $\epsilon$. The data for the correlation time, although not definitive,  suggest that there may be a sharp onset of SDC at $\epsilon _c > 0$. Beyond these heuristic considerations, we have only descriptive statistics to characterize the general features of the motion.  The higher moments of the structure factor which we have measured do not lend themselves to direct interpretation, but may be useful in the comparison of simulations to the data.  In general, SDC serves to highlight the difficult challenge that extensive chaos presents to our understanding of dynamical systems.

\acknowledgments 

We wish to thank Mike Cross, David Egolf, Henry Greenside, and Werner Pesch for interesting discussions on STC, and Robert Ecke, Yuchou Hu, Ning Li, and Steven Trainoff for discussions about structure factors and image analysis. This research was supported by the Department of Energy through Grant DE-FG03-87ER13738. S. W. M. acknowledges support from The Natural Sciences and Engineering Research Council of Canada, and E. B. from the Deutsche Forschungsgemeinschaft.

\begin{figure}
\caption{The sequence of patterns observed as $\epsilon$ is increased. (a): Nearly straight rolls at $\epsilon=0.043$. (b): Weakly time dependent ``frustrated" state  at $\epsilon=0.245$, just below the onset of SDC.}
\label{patternsequence_ab}
\end{figure}

\begin{figure}
\caption{The sequence of patterns observed as $\epsilon$ is increased. (a): At $\epsilon=0.536$, above the onset of SDC.  Spirals coexist with focus singularities at the edge of the cell. (b): At $\epsilon=0.894$, fully developed SDC fills the cell. }
\label{patternsequence_cd}
\end{figure}

\begin{figure}
\caption{The pattern observed at low $\epsilon=0.13$, with the sidewall heater turned on.  With the sidewall heater turned off, a pattern similar to Fig.  \protect{\ref{patternsequence_ab}}b is found instead. The heater has the effect of changing the roll orientation from perpendicular to parallel at the sidewall. }
\label{roundcellheaterloweps}
\end{figure}

\begin{figure}
\caption{The patterns in the SDC regime at $\epsilon=0.644$, (a) with the sidewall heater turned off, and (b) with the sidewall heater on.  SDC is insensitive to the sidewall conditions, except very near the edges.  }
\label{roundcellheaterchaos}
\end{figure}

\begin{figure}
\caption{Patterns observed in the square cell with sidewall heating applied to the upper and lower edges of the cell. (a) $\epsilon=0.17$, showing a ``frustrated" state, (b) SDC at $\epsilon=0.77$.  }
\label{sqcell}
\end{figure}

\vfill\eject

\vfill\eject

\vfill\eject

\begin{figure}
\caption{ A snapshot of the spacetime structure of fully developed SDC in the circular cell. (a) A space-space shadowgraph image at $\epsilon=1.032$.  The image is $106 d$ wide. (b) A spacetime slice of SDC, with the space coordinate horizontal and time running downward.  The image represents spacetime data along a horizontal line near the midpoint of image (a), spread out over a time interval of $363 t_v$.  The dimensions of (b) correspond to $46 \xi$ in the space direction and $55 \tau$ in time.}
%
%
\label{spacetime}
\end{figure}
\vfill\eject

\vfill\eject

\vfill\eject

\vfill\eject

\vfill\eject

\end{document}